\documentclass[twocolumn]{aastex631}
\usepackage{amsmath}
\usepackage{booktabs}
\usepackage{threeparttable}
\usepackage{CJK}

\begin{document}

\title{Sun-as-a-star spectroscopic observations of the line-of-sight velocity of a solar eruption on October 28, 2021}

\begin{CJK*}{UTF8}{gbsn}

\correspondingauthor{Hui Tian (田晖)}
\email{huitian@pku.edu.cn}

\author[0000-0002-7421-4701]{Yu Xu (徐昱)}
\affiliation{School of Earth and Space Sciences, Peking University, Beijing 100871, China}

\author[0000-0002-1369-1758]{Hui Tian (田晖)}
\affiliation{School of Earth and Space Sciences, Peking University, Beijing 100871, China}


\author[0000-0003-4804-5673]{Zhenyong Hou (侯振永)}
\affiliation{School of Earth and Space Sciences, Peking University, Beijing 100871, China}

\author[0000-0002-4973-0018]{Zihao Yang (杨子浩)}
\affiliation{School of Earth and Space Sciences, Peking University, Beijing 100871, China}

\author[0000-0002-6641-8034]{Yuhang Gao (高宇航)}
\affiliation{School of Earth and Space Sciences, Peking University, Beijing 100871, China}
\affiliation{Centre for mathematical Plasma Astrophysics (CmPA), KU Leuven, Celestijnenlaan 200B bus 2400, B-3001 Leuven, Belgium}

\author[0000-0003-2686-9153]{Xianyong Bai (白先勇)}
\affiliation{National Astronomical Observatories, Chinese Academy of Sciences, Beijing, 100012, China}

\begin{abstract}
The propagation direction and true velocity of a solar coronal mass ejection, which are among the most decisive factors for its geo-effectiveness, are difficult to determine through single-perspective imaging observations. Here we show that Sun-as-a-star spectroscopic observations, together with imaging observations, could allow us to solve this problem. Using observations of the Extreme-ultraviolet Variability Experiment onboard the \emph{Solar Dynamics Observatory}, we found clear blue-shifted secondary emission components in extreme ultraviolet spectral lines during a solar eruption on October 28, 2021. From simultaneous imaging observations, we found that the secondary components are caused by a mass ejection from the flare site. We estimated the line-of-sight (LOS) velocity of the ejecta from both the double Gaussian fitting method and the red-blue asymmetry analysis. The results of both methods agree well with each other, giving an average LOS velocity of the plasma of $\sim 423~\rm{km~s^{-1}}$. From the $304$ \AA~image series taken by the Extreme Ultraviolet Imager onboard the \emph{Solar Terrestrial Relation Observatory-A} (\emph{STEREO-A}) spacecraft, we estimated the plane-of-sky (POS) velocity from the \emph{STEREO-A} viewpoint {to be around $587~\rm{km~s^{-1}}$}. The full velocity of the bulk motion of the ejecta was then computed by combining the imaging and spectroscopic observations, which turns out {to be around $596~\rm{km~s^{-1}}$ with an angle of $42.4^\circ$ to the west of the Sun-Earth line and $16.0^\circ$ south to the ecliptic plane}. 
\end{abstract}

\section{introduction}
Coronal mass ejections (CMEs) are the largest-scale eruptive events on the Sun. Interaction of CMEs with the Earth magnetosphere could cause geomagnetic storms, which may pose major threats to man-made satellites and astronauts \citep{Gonzalez1994,Echer2008}. The propagation direction and true velocity are among the key factors that affect the geo-effectiveness of a CME \citep{Moon2005, Kim2008}, and thus their accurate determination is of great importance.

One way to obtain the true velocity of a CME is to combine its plane-of-sky (POS) velocity and its line-of-sight (LOS) velocity. The POS velocities of CMEs are often obtained from imaging observations with coronagraphs or extreme ultraviolet (EUV) imagers. The motion of ejected materials along the line of sight (LOS) towards observers can cause, according to the Doppler effect, blueshifts or blue-shifted components in spectral line profiles \citep[e.g.,][]{Tian2012,Tian2013}, through which one can calculate the LOS velocities of the ejecta. Thus, combining the POS and the LOS velocities, one can obtain the true velocities and propagation directions of CMEs. For example, \cite{Susino2013} studied a CME associated with a filament eruption and calculated the true velocities of both the CME front and the CME core. The POS velocities were obtained from images taken by the Large Angle and Spectrometric COronagraph (LASCO; \citealt{Brueckner1995}) onboard the \emph{Solar
and Heliospheric Observatory} (\emph{SOHO}) while the LOS velocities were computed using the blue-shifted component in O\,{\sc{vi}}\,$103.2~\rm{nm}$ \AA~line profiles observed by the slit spectrometer Ultraviolet Coronagraph Spectrometer (UVCS; \citealt{Kohl1995}) onboard \emph{SOHO}. The true velocities of the CME front and core were then computed using the geometric relationship, which turned out to be $\sim 578~\rm{km~s^{-1}}$ and $\sim 583~\rm{km~s^{-1}}$, respectively. 

Apart from estimating true velocities using the LOS and POS velocities, another method for reconstructing the three-dimensional (3D) propagation track of a CME is using multi-viewpoint observations. The twin \emph{Solar Terrestrial Relation Observatory} (\emph{STEREO};\citealt{Kaiser2008}) satellites provide view angles off the Sun-Earth line and offer opportunities for 3D reconstructions of CME tracks through multiple-viewpoint observations. Several 3D reconstruction techniques were developed, such as the tie-pointing method (e.g. \citealt{Inhester2006,Thompson2009,Liewer2011}), the polarization ratio technique \citep{Moran2004} and the mask fitting method \citep{Feng2012,Feng2013,Ying2022}. \cite{Mishra2013} estimated the true velocity of a CME using the data from COR2 in the Sun Earth Connection Coronal and Heliospheric Investigation (SECCHI;\citealt{Howard2008}), and predicted the arrival time of that CME by assuming that the velocity remains constant beyond the field-of-view (FOV) of the COR2. \cite{Susino2014} tracked the position and velocity variation of an erupting filament using the image series from the Extreme Ultraviolet Imager (EUVI;\citealt{Wulser2004}) based on the tie-pointing technique. They also reconstructed the 3D shape of the CME front from the polarization ratio method using the observations from SECCHI/COR1. By combining both the tie-point method and a supervised computer {vision} algorithm, \cite{Braga2017} obtained the velocity vectors of $17$ CME events using the imaging observations from the twin STEREO spacecraft.

The restriction of the first method (i.e. combining the POS and LOS velocities) in estimating true velocities mainly comes from the limited field-of-view (FOV) of a slit spectrometer, which significantly lowers the opportunity to capture CMEs. Also, this method can only be conducted when the slit covers the CME area during its early expansion. The second method (i.e. using multi-viewpoint observations) relies on observations from multiple viewpoints, which requires appropriate positions of different spacecraft to ensure simultaneous detection of the same single CME. This requirement often means a large increase in the cost of the whole mission. A low-cost Sun-as-a-star spectrograph that measures the full-disk integrated EUV spectra{, which is similar to that in stellar observations,} could overcome such drawbacks, as it is able to catch almost all CMEs originating from the front side of the solar disk and provide continuous monitoring of the early propagation of CMEs \citep{Yang2022}. Moreover, such observations could provide a unique reference for us to understand CMEs on other stars, whose signals can only be extracted from point-source spectra at present.

Current Sun-as-a-star spectral observations are conducted by the Extreme-ultraviolet Variability Experiment (EVE; \citealt{Woods2012}) onboard the \emph{Solar Dynamic Observatory} (\emph{SDO}; \citealt{Pesnell2012}). The Doppler shifts in spectra observed by EVE are still insufficiently investigated. Previous studies have revealed shifts of spectral line centroids during solar flares. For instance, 
\cite{Hudson2011} analyzed the EVE spectra during a flare and found redshifts around $17~\rm{km~s^{-1}}$ in the He\,{\sc{ii}}\,$30.4~\rm{nm}$ line as well as decreasing blueshifts in the Fe\,{\sc{xxiv}}\,$19.2~\rm{nm}$ line, which were interpreted as signatures of chromospheric evaporation. \cite{Brown2016} studied the Doppler shifts in hydrogen Lyman lines in six flares, calculated the Doppler velocities using three different methods and found that both redshifts and blueshifts in those Lyman lines were around $10~\rm{km~s^{-1}}$. They attributed the {blueshifts} in the low-temperature chromospheric lines to (1) the upflows of a cool, neutral hydrogen layer pushed upwards due to the heating in the deep chromosphere and (2) ejecta captured by the Atmospheric Imaging Assembly (AIA; \citealt{Lemen2012}) {onboard} \emph{SDO}. \cite{Chamberlin2016} investigated the shifts of the line centroids caused by instrumental effects and presented the optical correction function of wavelengths for the He\,{\sc{ii}}\,$30.4~\rm{nm}$ line. They also conducted a study of the X1.8 flare with a streamer blowout observed by \emph{SDO}/AIA in the $304$ \AA~passband on September 7, 2011. The peak Doppler velocity of the He\,{\sc{ii}}\,$30.4~\rm{nm}$ line calculated from the optically corrected spectra turned out to be $\sim 120~\rm{km~s^{-1}}$. \cite{Cheng2019} studied various types of plasma motions in two flare regions using spectra from EVE. Their single Gaussian fitting revealed blueshifts in the high-temperature lines and redshifts in the relatively cool lines in the flare gradual phase. They suggested that the blueshifts represent plasma upflows caused by chromospheric evaporation while the redshifts result from loop contraction or chromospheric condensation.

It is worth mentioning that a spectral line profile at the location of a solar eruption generally should consist of at least two components, a component from the nearly stationary background emission and the other one from the moving plasma \citep[e.g.,][]{Tian2012}. In the full-disk integrated spectra, the background component is normally much stronger. If the spectral resolution is too low, the total spectral profile may reveal a slight enhancement at the line wing or appear to be slightly shifted. In previous Sun-as-a-star spectroscopic investigations, a single Gaussian fitting was often used to reveal the Doppler shift of an entire line profile, which should be smaller than the true speed of the moving plasma \citep[e.g.,][]{Tian2011}. The blue-shifted secondary components caused by plasma motions along the LOS towards observers in the Sun-as-a-star spectra have not been {detected} yet. 

With recent EVE observations, we identified the blue-shifted secondary components in full-disk integrated spectra caused by a mass ejection and calculated the LOS velocity of the ejecta. We combined {this} with the plane-of-sky (POS) velocity from the \emph{STEREO-A} viewpoint to obtain {the full velocity (i.e. true velocity) of the bulk motion} of the ejected plasma. Section \ref{sec:observations} gives a brief introduction to the observations from \emph{SDO}/EVE and \emph{STEREO-A}/EUVI. Section \ref{sec:results} presents the data analysis results and relevant discussions. Finally,  we  summarize our finding in Section \ref{sec:conclusion}.\\

\section{observations}\label{sec:observations}
\indent The EVE onboard \emph{SDO} is a full-disk integrated spectrometer measuring the solar EUV irradiance from $6~\rm{nm}$ to $106~\rm{nm}$ with a spectral resolution of $\sim0.1~\rm{nm}$ and a time cadence of $\sim10~\rm{s}$. The sampling wavelength interval is $0.02~\rm{nm}$, which enables detection of Doppler velocities at tens of kilometers per second \citep{Hudson2011}. The Multiple EUV Grating Spectrographs-A (MEGS-A) included in EVE experienced an anomaly on May 26, 2014, and after that no spectra at wavelengths shorter than $33~\rm{nm}$ are available. The integration time for MEGS-B was then adjusted to $1$ minute for a higher signal-to-noise ratio. \\
\begin{figure*}
    \centering
    \includegraphics[width=0.8\textwidth]{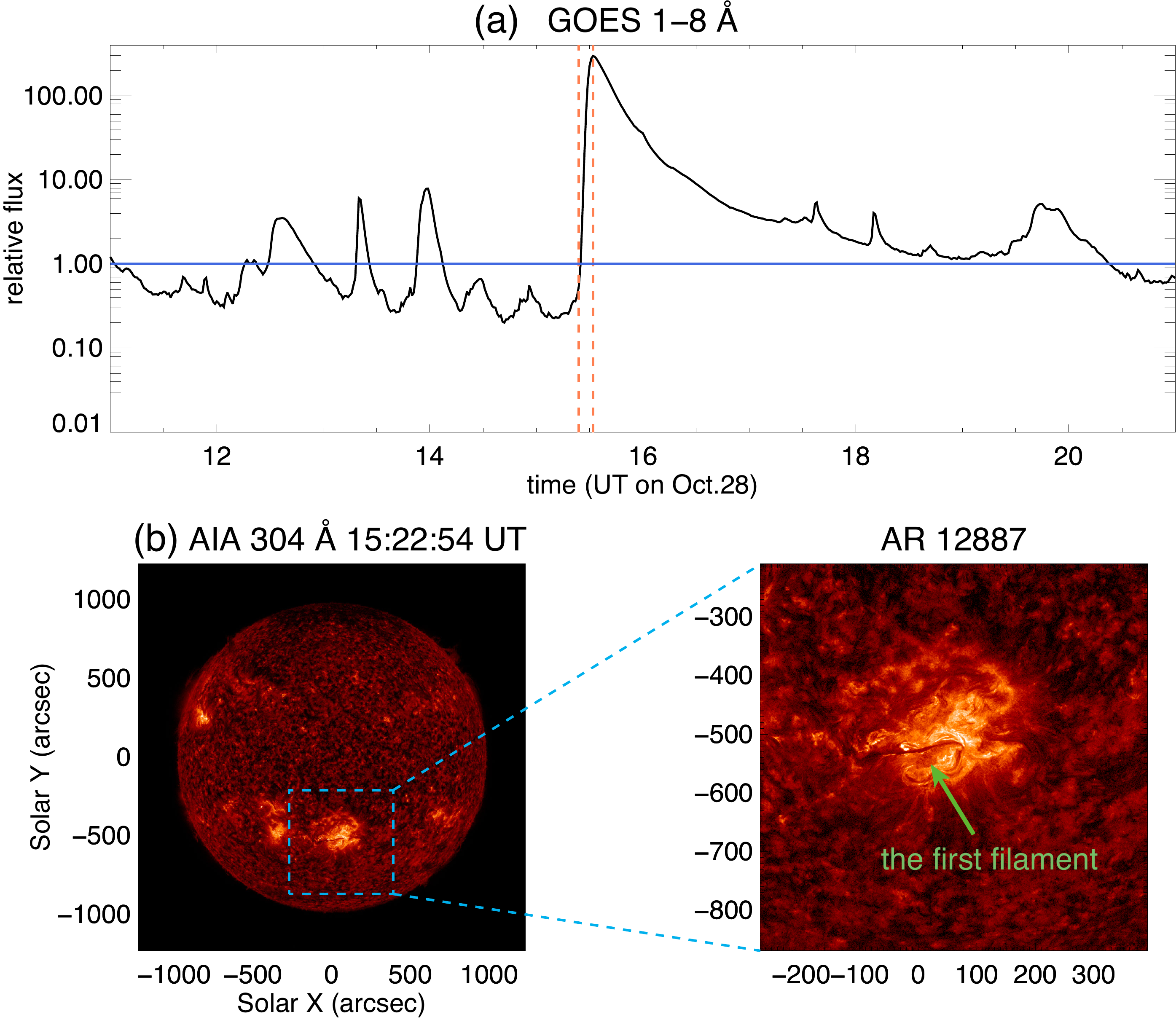}
    \caption{(a) The relative flux in $1-8$~\AA~from GOES. The red vertical dashed lines indicate the flare onset and peak at 15:24 UT and 15:32 UT, respectively. The blue line shows the quiescent level of X-ray flux. (b) The AIA $304$~\AA~image before the flare onset. The green arrow shows the location of the first filament.}
    \label{fig:full304xrt}
\end{figure*}
\indent The target mass ejection occurred on October 28, after an X1.0 flare in AR 12887. The flare onset is at 15:24 UT and the X-ray flux in $1-8$~\AA~measured by GOES-16 reaches its peak at 15:32 UT (see Fig. \ref{fig:full304xrt}(a)). A filament in the active region (AR) rises and appears to be heated to higher temperatures, becoming a bright structure in AIA 304~\AA~that erupts during the flare impulsive phase. It is then followed by another sympathetic filament eruption at around 15:35 UT, which is located at the east footpoint of the aforementioned filament. This event generates a global EUV wave, which was recently analyzed by \cite{Hou2022}. In Fig. \ref{fig:full304xrt}(b), the $304$~\AA~image taken by \emph{SDO}/AIA shows the location of the source AR (i.e. AR 12887) and that of the first filament whose rising and heating lead to the target ejection.\\
\begin{figure*}
    \centering
    \includegraphics[width=0.8\textwidth]{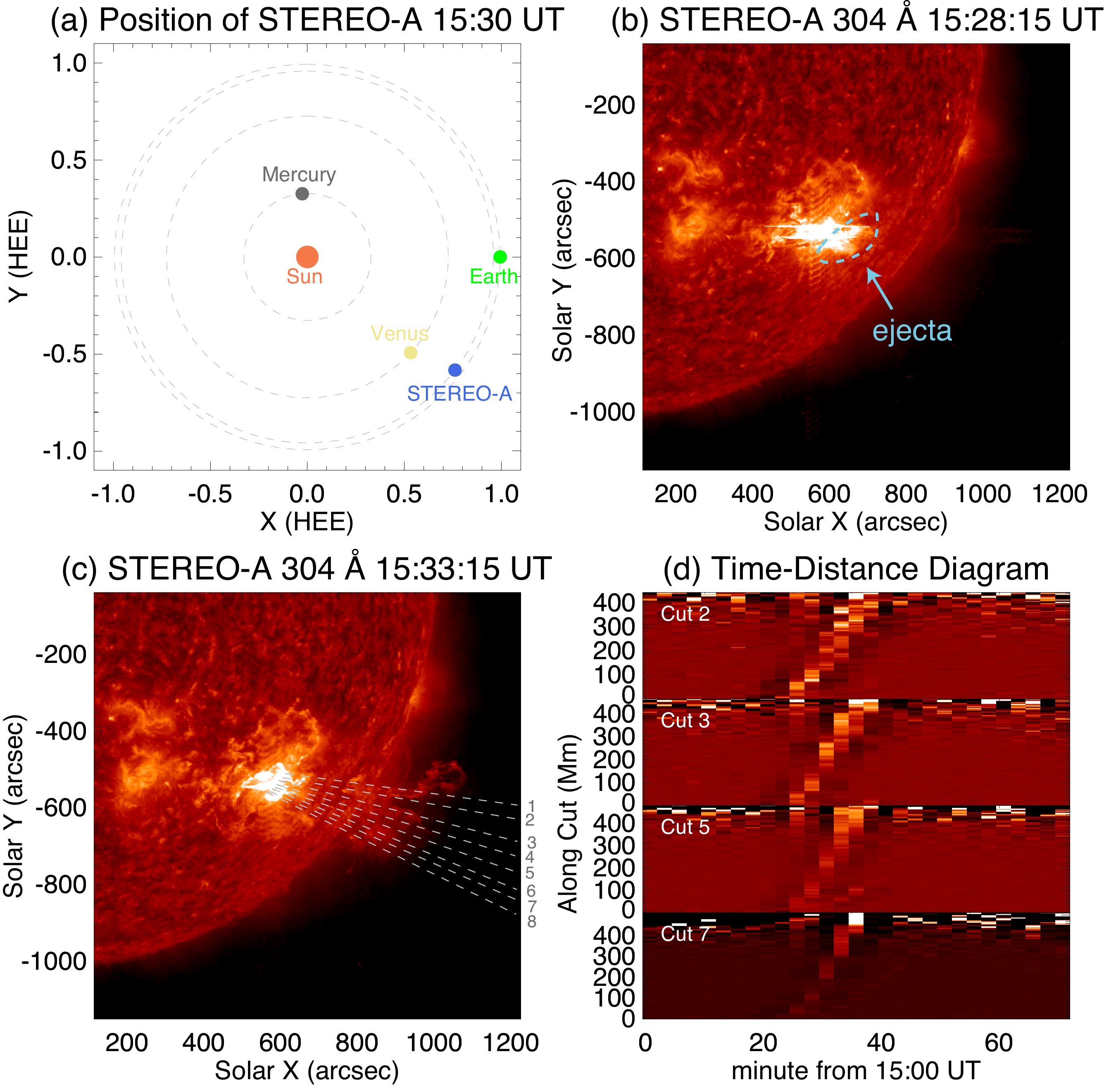}
    \caption{(a) The positions of \emph{STEREO-A} and the Earth at 15:30 UT on 2021 October 28 in the HEE coordinate system. (b) The ejecta observed by \emph{STEREO-A} in the $304$~\AA~passband at 15:28:15 UT. The blue dashed line outlines the bright ejecta. (c) The ejecta observed by \emph{STEREO-A} in the $304$~\AA~passband at 15:33:15 UT. The grey dashed lines are the cuts chosen for the time-space analysis (see the text in Sec. \ref{sec:velocity}) and the numbers at the end of the dashed lines label the cuts. (d) The time-distance diagrams for Cut $2$, Cut $3$, Cut $5$, and {Cut $7$} (see Sec. \ref{sec:velocity} for more details). }
    \label{fig:stereoa}
\end{figure*}
\indent In the meantime, \emph{STEREO-A}/EUVI captured the eruption at around $37.5^\circ$ east to the Sun-Earth line (see Fig. \ref{fig:stereoa}(b)(c)), and the positions of \emph{STEREO-A} and the Earth are shown in Fig. \ref{fig:stereoa}(a) under the Heliocentric Earth Ecliptic (HEE) coordinate system. The X-axis of the HEE system points from the center of the Sun towards the Earth and the Z-axis points along the ecliptic north pole \citep{Thompson2006}. The \emph{STEREO-A}/EUVI images the Sun in four channels (i.e. $30.4~\rm{nm}$, $17.1~\rm{nm}$, $19.5~\rm{nm}$, and $28.4~\rm{nm}$) spanning the $0.1$ to $20~\rm{MK}$ temperature range. We used the image series of $30.4~\rm{nm}$ for velocity analysis (see Sec. \ref{sec:velocity}). The time interval between two adjacent $30.4~\rm{nm}$ images is $2.5$ min.\\

\section{analysis and results}\label{sec:results}
\subsection{Velocity estimation from STEREO observations}\label{sec:velocity}
\indent We derived the POS (from the \emph{STEREO-A} viewpoint) velocity of the ejecta using a series of $304$~\AA~images from \emph{STEREO-A} observations. {The EUVI images were rectified using \emph{secchi\_prep.pro} in SolarSoftWare\footnote{https://www.lmsal.com/solarsoft} package with the keyword \emph{rotated\_on} applied to rotate the solar north up in the images. We chose eight cuts (grey dashed lines in Fig. \ref{fig:stereoa}(c)) along the directions of plasma motion. Then we derived the speeds of the ejecta along the eight directions from time-distance diagrams. The measurements are under the helioprojective-cartesian (HPC) coordinate system in EUVI images. The HPC system owns a Z-axis along the Sun-observer line and Y-axis perpendicular to Z-axis in the plane containing both the solar north pole and the Z-axis. The X-axis is perpendicular to both Y-axis and Z-axis, forming a right-handed coordinate system (see \citealt{Thompson2006} for more details). \\} 
\indent No obvious acceleration was seen in the time-distance diagrams (Fig. \ref{fig:stereoa}(d)), which can be attributed to the poor time cadence of EUVI as the initial acceleration process of the mass eruption was completed within $2$ minutes. We then reasonably assumed that the ejecta moved along each cut at a constant speed. At each time step between 15:20 UT and 15:50 UT along each cut, the location of the moving plasma front was identified as the position where the intensity reaches its maximum value. For each cut, locations of the fronts form a track of the ejecta and we applied a linear fitting to the track to estimate the speed. The average velocity of the ejecta $v_{\rm{POS}}$ {and its angle with respect to the $+$X-axis direction of the HPC system (i.e. Solar X in Fig. \ref{fig:stereoa}(b)--(c))} were then computed using the following equations,
\begin{equation}
v_{\rm{POS}}=\frac{\sum_{i=1}^{8} v_i P_i}{\sum_{i=1}^{8} P_i}
\label{vpos}
\end{equation}
\begin{equation}
\theta=\frac{\sum_{i=1}^{8} \theta_i P_i}{\sum_{i=1}^{8} P_i}
\label{theta}
\end{equation}
{where $v_i$ and $\theta_{i}$ is the speed along Cut $i$ ($i=1,2,..,8$) and the angle between the $+$X-axis direction of the HPC system and the direction of Cut $i$, respectively. All angles are measured counterclockwise {with respect to the +X-axis direction of the HPC system}. $P_i$ is the Pearson correlation coefficient that indicates the linear correlation between the front locations and the corresponding times. The identification of the ejecta fronts by maximum intensity values in the time-distance diagrams occasionally results in non-linearity in the track of the ejecta, so we used $P_i$ as weighting factors in Eq. \eqref{vpos}--\eqref{theta}. The average POS velocity turned out to be around $587~\rm{km~s^{-1}}$ with an angle of around $-17.8^\circ$ to the $+$X-axis direction of the HPC system.}\\

\subsection{EVE observations of spectral line intensities}
\begin{figure*}
    \centering
    \includegraphics[width=0.9\textwidth]{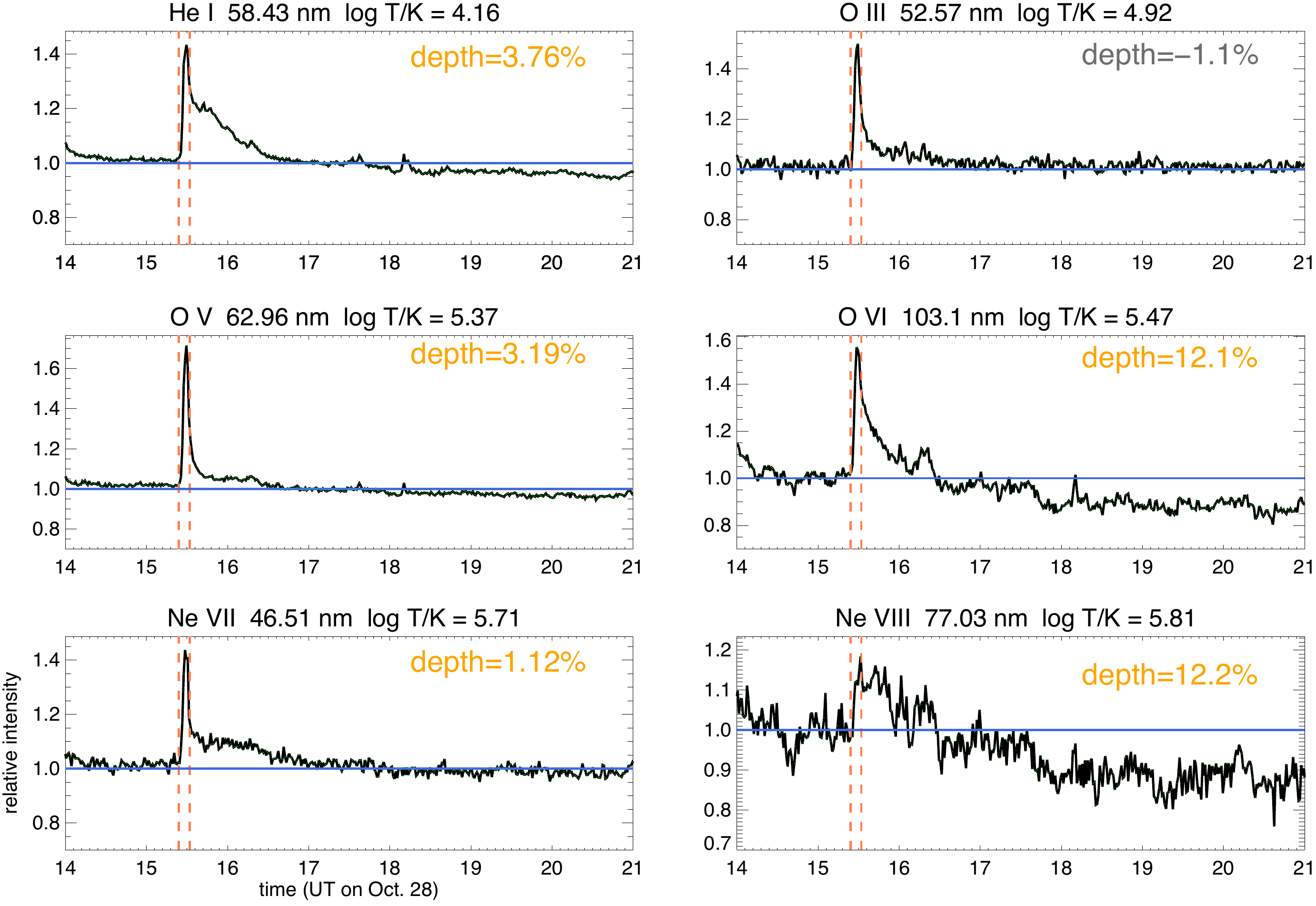}
    \caption{The relative intensity variations for the selected spectral lines. The red vertical dash lines indicate the flare onset and peak at 15:24 UT and 15:32 UT, respectively. The solid blue line represents the quiescent intensity level. The average dimming depth during 19:00 UT to 21:00 UT for each line is marked in each panel (positive for dimming while negative for enhancement).}
    \label{fig:irradiance}
\end{figure*}

We chose the spectra observed by \emph{SDO}/EVE MEGS-B from 11:00 UT to 21:00 UT for analysis. First, we selected six spectral lines (listed in Table \ref{tab:lines}) for detailed analysis and checked line blends for each of them using the CHIANTI 10.0.1 {\citep{Dere1997,DelZanna2021}} database. Lines listed in Table \ref{tab:lines} do not have obvious blends except for a red wing blend in Ne\,{\sc{vii}}\,$46.52~\rm{nm}$ with Ca\,{\sc{ix}}\,$46.63~\rm{nm}$. Second, we calculated the intensity variation with time for each line. We applied a single Gaussian fitting (linear background for the continuum) to all the lines in Table \ref{tab:lines} at each time step and obtained their line intensities. We defined a \emph{quiescent time}, which was chosen to start from 11:00 UT to 13:00 UT when no obvious flare is observed (except for a small C2.2 flare at around 12:26 UT). The quiescent intensity level for each line was obtained by averaging the line intensities over the \emph{quiescent time} period. The intensity variations for each line relative to their quiescent levels are displayed in Fig. \ref{fig:irradiance}. The pre-flare relative intensities of all lines maintain a level around $1$, which validates the choice of the \emph{quiescent time} to be reasonable. The intensities of all lines begin to rise right after the flare onset. The lines formed below $10^{5.8}~\rm{K}$ reach their peaks before the flare peak, and the Ne\,{\sc{viii}}\,$77.03~\rm{nm}$ line with a formation temperature above $10^{5.8}~\rm{K}$ reaches {its peak around the time of the GOES soft X-ray peak}. We noticed that almost all the spectral lines show dimming signatures (i.e. intensity dropping below the quiescent level) during and after the flare gradual phase except for O\,{\sc{iii}}\,$52.57~\rm{nm}$. We roughly quantified the dimming depth by averaging the decrease of the relative intensity from 19:00 UT to 21:00 UT when the flare almost ends. The dimming depths {are} marked in Fig. \ref{fig:irradiance}, with positive values indicating dimming while negative enhancement. We suggest that the dimming is caused by the loss of plasma due to the mass ejection, in which the decrease of plasma density in the source region leads to the decrease of EUV emission. Signatures of CME-induced dimmings have been identified from EVE observations \citep[e.g.,][]{Mason2014,Mason2016,Harra2016}. In these previous observations, the dimmed lines usually have a minimum formation temperature of $\sim 10^{5.8}~\rm{K}$. Here we show that the dimming also exists in relatively cooler lines such as He\,{\sc{i}}\,$58.43~\rm{nm}$, O\,{\sc{v}}\,$62.96~\rm{nm}$ and O\,{\sc{vi}}\,$103.1~\rm{nm}$, which suggests that some plasma in the lower atmosphere rises and erupts, resulting in the mass loss dimming.\\
\begin{table}
    \caption{Spectral lines selected for analysis in EVE MEGS-B dataset}
    \label{tab:lines}
    \centering
    \begin{threeparttable}
        \begin{tabular}{*4{c}}
            \toprule
             Ion 
             & Rest Wavelength\tnote{a} (nm)& log (T/K) \\
            \midrule
            He I   
            & 58.4303 & 4.16\\
             O III  
             & 52.5770 & 4.92\\
             O V    
             & 62.9683 & 5.37\\
             O VI   
            & 103.191 & 5.47\\
             Ne VII\tnote{b} 
             & 46.5220 & 5.71\\
             Ne VIII
             & 77.0365 & 5.81\\
            \bottomrule
        \end{tabular}
        \begin{tablenotes}
            \footnotesize
            \item[a] The rest wavelengths were obtained by averaging the centroid locations of line profiles observed during the \emph{quiescent time} (see the text for details).
            \item[b] The red wing of Ne\,{\sc{vii}}\,$46.52~\rm{nm}$ is contaminated by Ca IX $46.63~\rm{nm}$. 
        \end{tablenotes}
    \end{threeparttable}
\end{table}

\subsection{EVE observations of the LOS velocities}\label{sec:losvelo}
\begin{figure*}
    \centering
    \includegraphics[width=\textwidth]{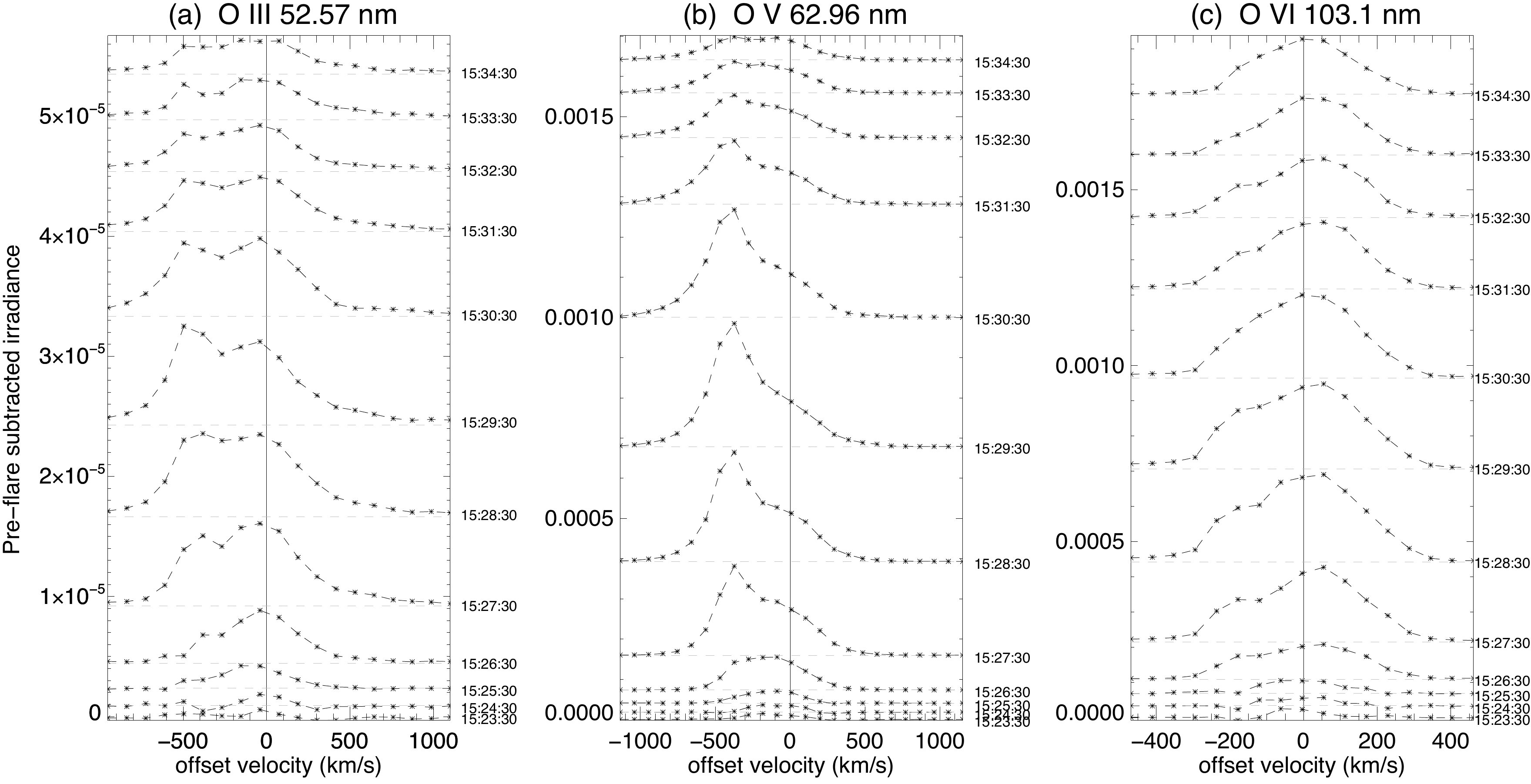}
    \caption{The pre-flare subtracted line profiles for O\,{\sc{iii}}\,$52.57~\rm{nm}$ (a), O\,{\sc{v}}\,$62.96~\rm{nm}$ (b), and O\,{\sc{vi}}\,$103.1~\rm{nm}$ (c) from 15:23:30 UT to 15:34:30 UT. The observation times are labeled on the right side of each profile and the y-axis is offset for a better illustration. The grey dashed lines indicate the level of spectral irradiance during the quiescent time. The vertical black lines indicate the rest wavelengths. The negative offset velocities represent the blue wings while the positive the red.}
    \label{fig:flareexcess}
\end{figure*}

\begin{figure}
    \centering
    \includegraphics[width=0.4\textwidth]{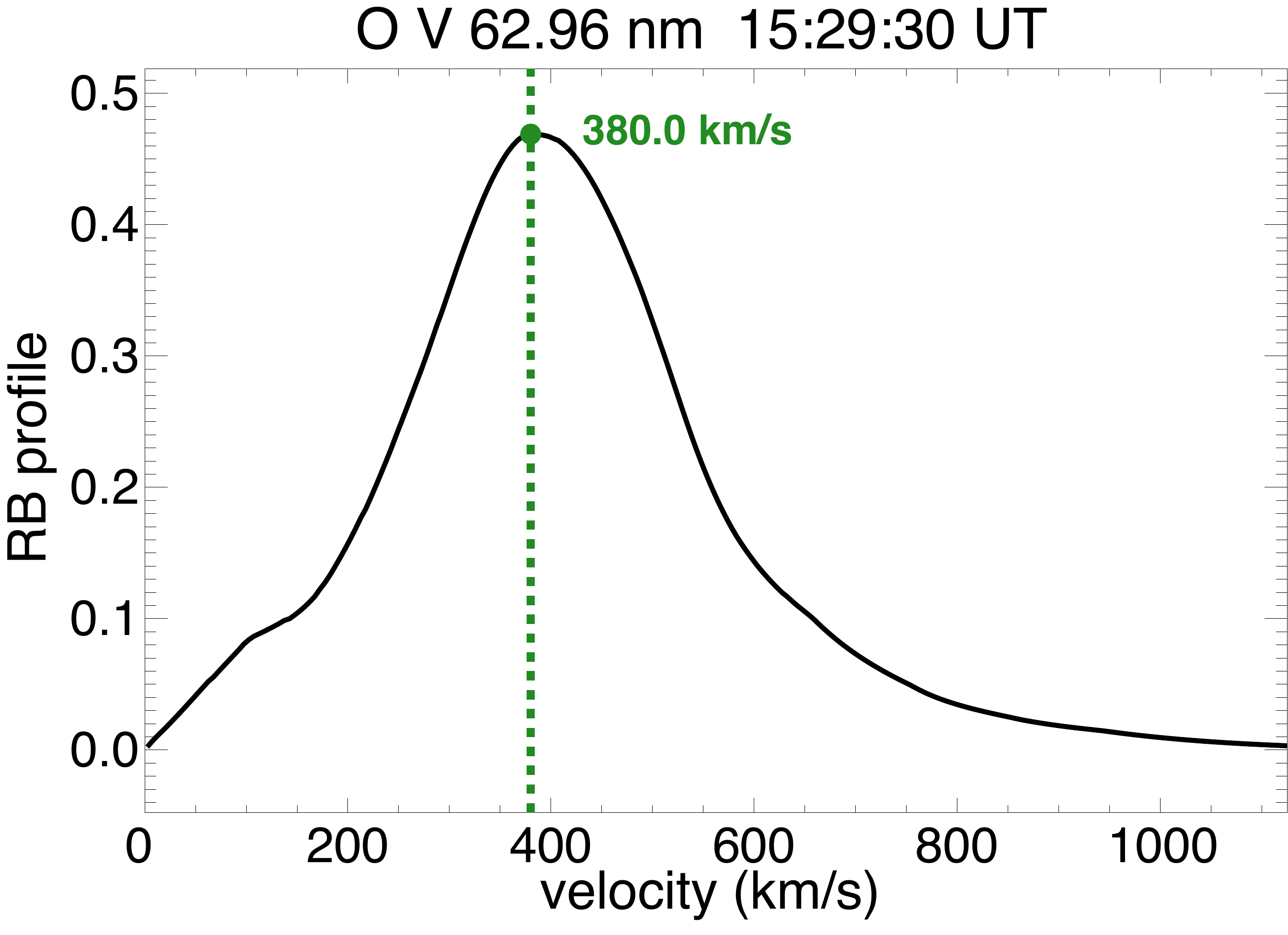}
    \caption{The RB profile of the line {O\,{\sc{v}}\,$62.96~\rm{nm}$} at 15:29:30 UT. The green vertical dashed line indicates the velocity corresponding to the peak of the RB profile.}
    \label{fig:rbp}
\end{figure}

\begin{figure*}
\begin{interactive}{js}{velos3d.zip}
    \centering
    \includegraphics[width=0.9\textwidth]{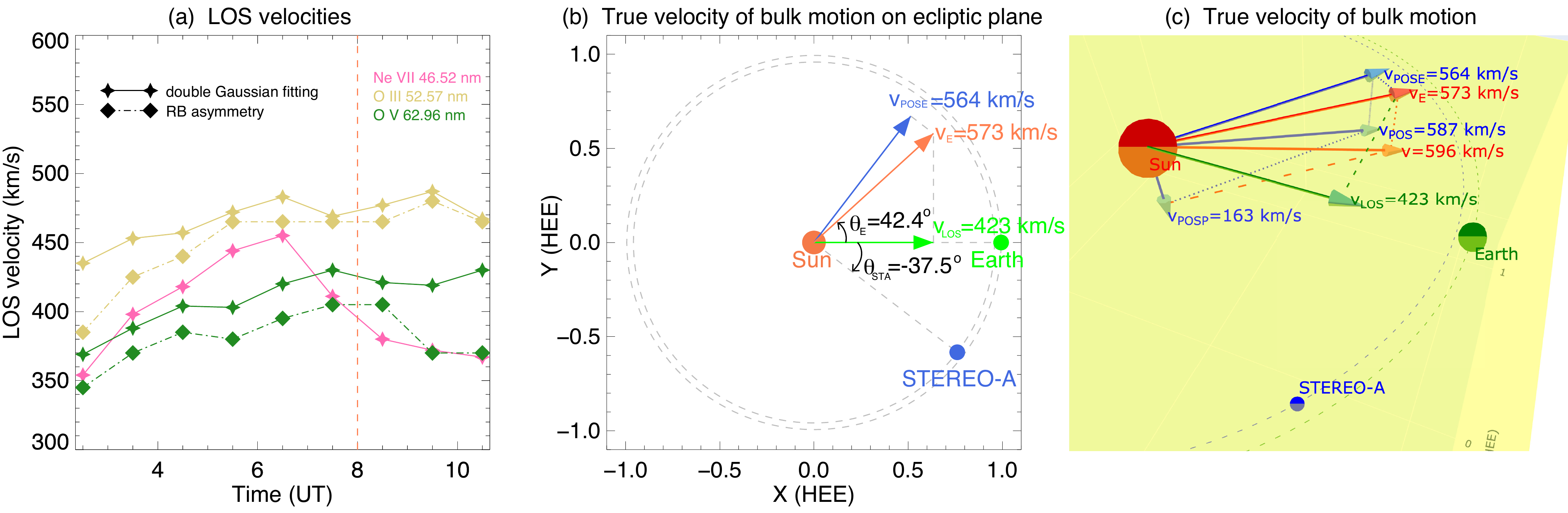}
\end{interactive}
\caption{(a) LOS velocities calculated from the double Gaussian fitting method (stars) and from the RB asymmetry analysis (diamonds) for O\,{\sc{iii}}\,$52.57~\rm{nm}$ (yellow), O\,{\sc{v}}\,$62.96~\rm{nm}$ (green), and Ne\,{\sc{vii}}\,$46.52~\rm{nm}$ (pink). The vertical red dashed line indicates the time of the flare peak. (b) A sketch of the {full velocity} calculation on the ecliptic plane for the mass ejection. {(c) A snapshot of the full velocity calculation for the ejecta in 3D space. The yellow plane represents the ecliptic plane. The red/green/blue sphere shows the location of the Sun/Earth/STEREO-A. The arrows indicates the velocity vectors and the corresponding speed is labeled at the header of each arrow. A 3D interactive version of panel (c) is available online and is created by \emph{Plotly} (https://plotly.com). One can zoom in and rotate the 3D plot using the toolbar located at the upper right corner of the figure. The interactive version provides a better understanding of the direction of the bulk motion of the ejecta in 3D space.}}
    \label{fig:velos}
\end{figure*}

\indent We examined the line profile variations for each line and identified emission enhancement in the blue wing during the impulsive phase and after the flare peak. For each spectral line, we obtained a reference line profile by averaging the line profiles during the \emph{quiescent time}, and then subtracted the reference line profile from the observed line profiles to obtain the pre-flare subtracted line profiles. Temporal evolutions of the pre-flare subtracted line profiles of O\,{\sc{iii}}\,$52.57~\rm{nm}$, O\,{\sc{v}}\,$62.96~\rm{nm}$, and {O\,{\sc{vi}}\,$103.1~\rm{nm}$} are displayed in Fig. \ref{fig:flareexcess}. In order to diminish the effect of the sympathetic filament eruption at around 15:35 UT (as described in Sec. \ref{sec:observations}), we only display the pre-flare subtraction analysis results from 15:23:30 UT to 15:34:30 UT. The wavelength $\lambda$ was converted into the offset velocity $v_c$ through the following equation,
\begin{equation}
v_c=\frac{\lambda-\lambda_0}{\lambda_0}c
\label{vc}
\end{equation}
where $c=2.9979\times 10^{5}~\rm{km~s^{-1}}$ is the light speed in the vacuum. $\lambda_0$ is the rest wavelength of the line, which was computed by averaging the central wavelengths, obtained from the single Gaussian fitting, during the \emph{quiescent time} period. Figure \ref{fig:flareexcess} shows that the enhancement peaks near the rest wavelength and in the blue wing at around $500~\rm{km~s^{-1}}$ in each of the spectral lines, which we suggested was caused by the outward movement of the heated plasma. The pre-flare subtraction was also conducted for the other three lines and the variations in He\,{\sc{i}}\,$58.43~\rm{nm}$ and Ne\,{\sc{vii}}\,$46.52~\rm{nm}$ are similar to those in the oxygen lines. {The emission enhancements at the blue wings of those lines suggest that the heated filament contains plasma with a temperature range of about 0.02 MK (i.e. the formation temperature of He\,{\sc{i}}\,$58.43~\rm{nm}$) -- 0.5 MK (i.e. the formation temperature of Ne\,{\sc{vii}}\,$46.52~\rm{nm}$).} However, the variations in the Ne\,{\sc{viii}}\,$77.03~\rm{nm}$ line are much more fluctuated than the others and more than one peak were detected. The existence of fluctuation in the Ne\,{\sc{viii}}\,$77.03~\rm{nm}$ line still needs further investigation.\\
\begin{figure*}
    \centering
    \includegraphics[width=\textwidth]{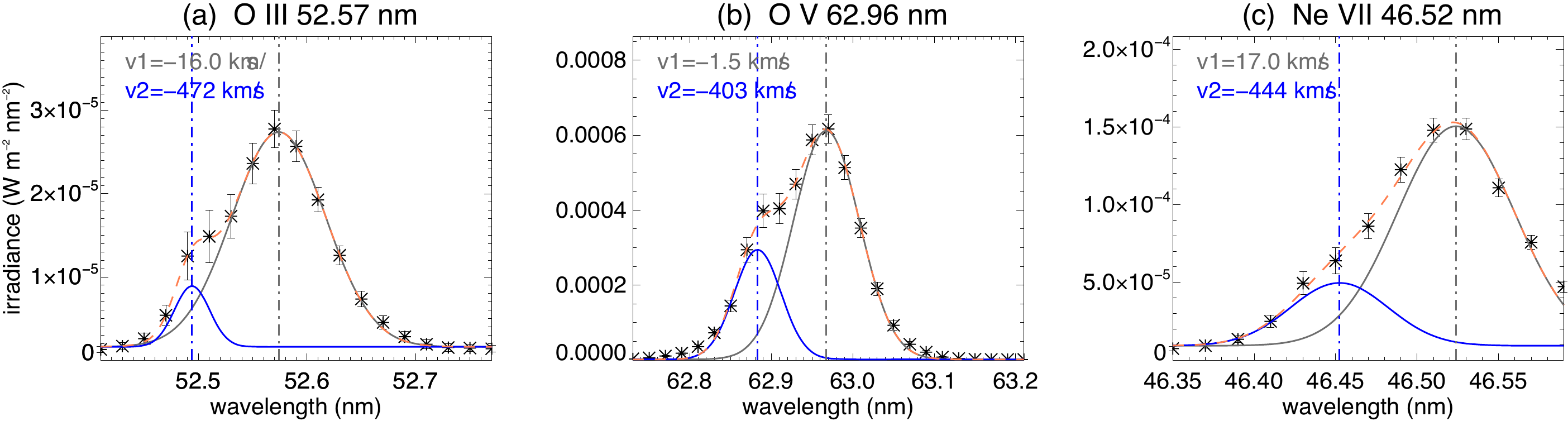}
    \caption{Line profiles of O\,{\sc{iii}}\,$52.57~\rm{nm}$ (a), O\,{\sc{v}}\,$62.96~\rm{nm}$ (b), and Ne\,{\sc{vii}}\,$46.52~\rm{nm}$ (c) observed at 15:29:30 UT and the results of double Gaussian fitting. {The stars are the observed spectral irradiances while the error bars are derived from the measurement precision contained in the EVE observation files.} The red dashed lines are the results of the double Gaussian fitting, and the grey curves represent the primary component while the blue the secondary. The vertical lines indicate the locations of the center of the primary (grey) and that of the secondary (blue) components. The Doppler velocities derived from the double Gaussian fitting are labeled as $v1$ and $v2$ for the primary and secondary components, respectively.}
    \label{fig:2ndcomp}
\end{figure*}
\indent In order to further investigate the LOS velocity of the moving plasma, we analyzed the observed spectral profiles of the six emission lines during the flare. We found obvious secondary components in the blue wings of O\,{\sc{iii}}\,$52.57~\rm{nm}$, O\,{\sc{v}}\,$62.96~\rm{nm}$, and Ne\,{\sc{vii}}\,$46.52~\rm{nm}$ during the impulsive phase and {shortly} after the flare peak. The spectral profiles of these lines at a specific time (15:29:30 UT) are shown in Fig. \ref{fig:2ndcomp}. The small bulges in the blue wings suggest the presence of secondary components. To the best of our knowledge, this is the first time that a distinct secondary emission component caused by a mass ejection is detected in the Sun-as-a-star spectra. Through a single Gaussian fitting, previous investigations of several flares (see e.g., \citealt{Hudson2011,Brown2016,Chamberlin2016,Cheng2019}) all found shifts of line centroids that are mostly smaller than $100~\rm{km~s^{-1}}$. In these observations, no obvious secondary components were seen. The obvious bulges in the blue wings of the spectral lines (as shown in Fig. \ref{fig:2ndcomp}) facilitate the conduction of double Gaussian fitting to derive the LOS speed of the ejecta.\\
\indent We conducted the double Gaussian fitting for line profiles of O\,{\sc{iii}}\,$52.57~\rm{nm}$, O\,{\sc{v}}\,$62.96~\rm{nm}$, and Ne\,{\sc{vii}}\,$46.52~\rm{nm}$ observed during the impulsive phase and several minutes after the flare peak. We adopted the following expression for the double Gaussian fitting:
\begin{equation}
I_{fit}(\lambda)=b_0+i_{1}e^{-\frac{(\lambda-\lambda_1)^2}{w_1^2}}+i_{2}e^{-\frac{(\lambda-\lambda_2)^2}{w_2^2}}
\end{equation}
where $b_0$ was a constant background representing the continuum. $i_{n}$ and $\lambda_{n}$ ($n=1,2$) were the line {peak} intensity and central wavelength of the $n$th component, respectively. We followed the definition of exponential width in \cite{Tian2012} and used $w_{n}$ ($n=1,2$) to represent the exponential width (width hereafter) of the $n$th component. We regarded the primary component as the $1$st component, which is located near the rest wavelength and represents emission from the background solar atmosphere. The secondary (or $2$nd) component is located in the blue wing of the line, which results from the outward moving ejecta. {We adopted the double Gaussian fitting procedure used by \cite{Tian2011}. The procedure requires initial guess values of centroids, widths, and peak intensities of the two Gaussian components and their variation ranges as inputs and iterates until a global minimization of the difference between the observed spectrum and the fitted one is reached.} The initial values of the parameters were set as follows. We took the minimum spectral irradiance in an observed line profile as the initial constant background $b_0$ and allowed it to vary from $75\%$ to $125\%$ of the initial value during the iterations. The initial central wavelengths of the primary and secondary components were set to be the rest wavelength and the wavelength corresponding to an offset velocity of $-400~\rm{km~s^{-1}}$, respectively. The central wavelength of the primary component was allowed to vary from $-20~\rm{km~s^{-1}}$ to $20~\rm{km~s^{-1}}$, a rough range of the shifts of line centroids derived from single Gaussian fitting found in previous EVE observations \citep{Hudson2011,Brown2016}. The centroid of the secondary component was allowed to vary within the blue wing. The initial peak intensity values of the primary and secondary components were set to be the spectral irradiances at the offset velocities of zero and $-400~\rm{km~s^{-1}}$, respectively. The peak intensity of the primary component was allowed to vary within $[75\%,125\%]$ of the initial value, and that of the secondary component was allowed to vary from $0$ to the maximum spectral irradiance of the line profile. The initial widths of both components were set to equal the average width during the \emph{quiescent time}. And their variation ranges during the iterations were $[50\%,150\%]$ and $[50\%,125\%]$ of the initial value, respectively. All the initial value settings were the results of trial and error. The uncertainty of the double Gaussian fitting coming from a large number of free parameters causes the failure of fitting sometimes. We selected the valid fitting results based on the principle that the fitted profiles are located within the measurement error ranges of the observed profiles.\\
\begin{figure*}
    \centering
    \includegraphics[width=0.9\textwidth]{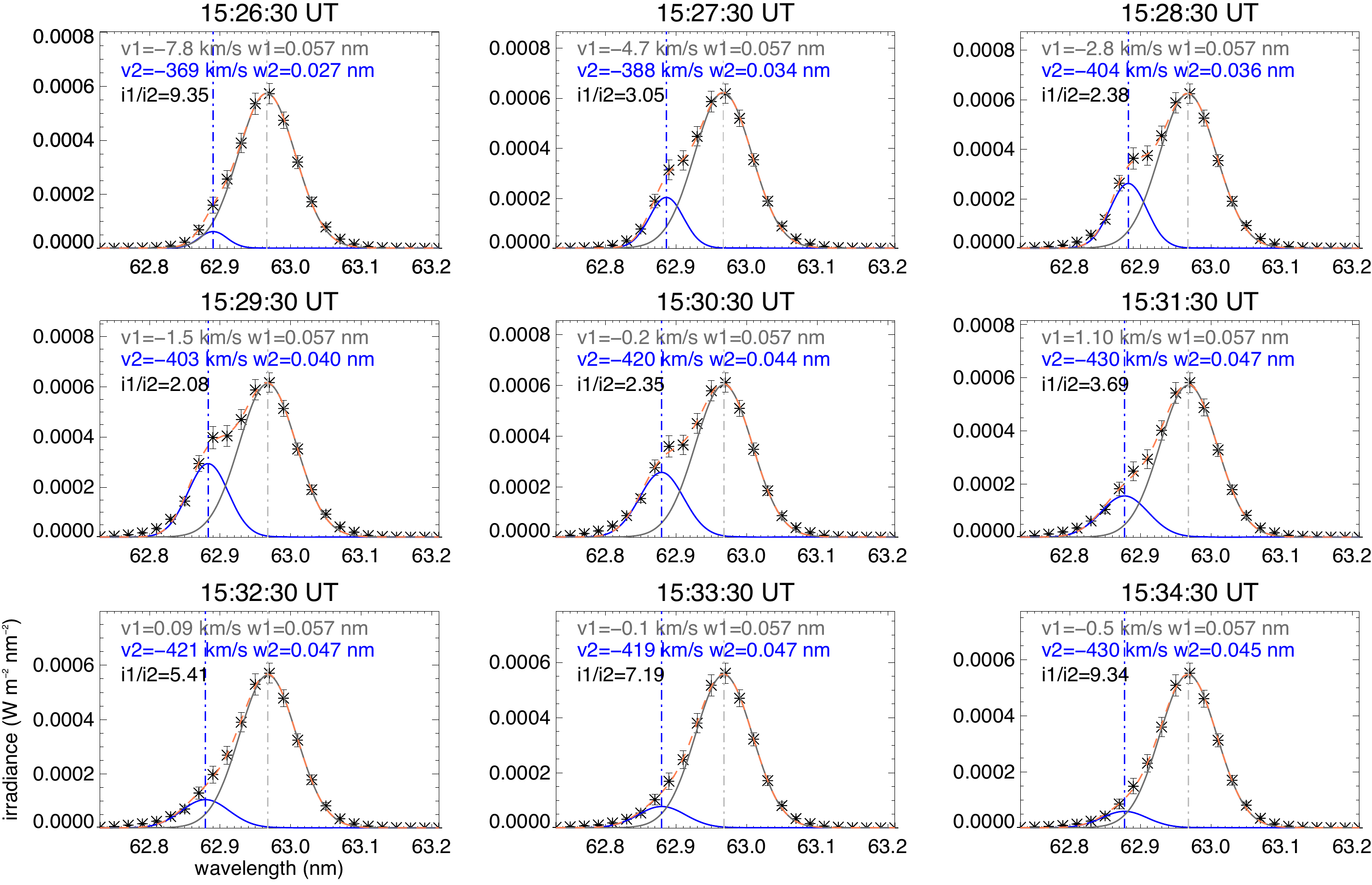}
    \caption{The double Gaussian fitting results for O\,{\sc{v}}\,$62.96~\rm{nm}$ from 15:26:30 UT to 15:34:30 UT. The line styles, colors, stars and error bars are the same as those in Fig. \ref{fig:2ndcomp}. The Doppler velocity $v_{n}$ ($n=1,2$) and width $w_n$ ($n=1,2$) of the two components are labeled in each panel, where the grey (blue) represents the parameters of the primary (secondary) component. The {peak} intensity ratio of the primary component to the secondary component ($i1/i2$) is also shown in each panel.}
    \label{fig:ov}
\end{figure*}
\indent The double Gaussian fitting results at 15:29:30 UT for the three lines with obvious secondary components (i.e. O\,{\sc{iii}}\,$52.57~\rm{nm}$, O\,{\sc{v}}\,$62.96~\rm{nm}$, and Ne\,{\sc{vii}}\,$46.52~\rm{nm}$) are shown in Fig. \ref{fig:2ndcomp}, {where we interpolated the wavelength points to make the fitted curves look much smoother.} All three lines reveal a high-speed component with a velocity of $400$--$500~\rm{km~s^{-1}}$. The fitting results for O\,{\sc{v}}\,$62.96~\rm{nm}$ at different time steps are shown in Fig. \ref{fig:ov}. The double Gaussian fitting failed in the first two minutes after the flare onset, so we only show the results from 15:26:30 UT to 15:34:30 UT. Our analysis indicates that the plasma at around $0.2~\rm{MK}$ (the formation temperature of the O\,{\sc{v}}\,$62.96~\rm{nm}$ line) moves at a speed around $400~\rm{km~s^{-1}}$ during the flare impulsive phase and {shortly} after the flare peak. The width of the primary component remains relatively constant, whereas that of the secondary component generally increases with time. This could be attributed to the expansion of the ejecta, which causes a wide velocity distribution and then results in a broadening of the line profiles. The variation of the {peak} intensity ratio of the two components shows that the secondary component reaches its peak at around 15:29:30 UT, which is $\sim 3$ minutes before the flare peak. The intensity ratio then gradually decreases. The velocities of the secondary components in all the three lines are plotted in Fig. \ref{fig:velos}(a). We can see a slight increasing trend of the speeds from 15:26 UT to 15:30 UT in all the three lines, likely indicating the continuous acceleration process of the ejecta.\\
\indent Another method for estimating the velocity of the secondary component is to use the red-blue (RB) asymmetry of the line profile (see e.g., \citealt{DePontieu2009,Tian2011}). The RB asymmetry profile (hereafter, RB profile) was calculated from the following equation
\begin{equation}
RB(v_c) =\int_{-v_c-\delta v/2}^{-v_c+\delta v/2} I(v) dv-\int_{v_c-\delta v/2}^{v_c+\delta v/2} I(v)dv
\label{rbvc}
\end{equation}
where $v_c$ is the offset velocity computed using Eq. \eqref{vc} (here positive only) and $\delta v$ is the velocity step in the RB profile calculation. $I(v)$ is the observational or interpolated spectral irradiance around $v_c$. From the peak of the $RB(v_c)$ profile, we can infer parameters of the secondary component. We interpolated an observational line profile and adopted $\delta v=5~\rm{km~s^{-1}}$ to obtain the RB profile at each time step. We regarded the velocity where the RB profile peaks as the LOS velocity of the ejecta. Figure \ref{fig:rbp} shows the RB profile of the O\,{\sc{v}}\,$62.96~\rm{nm}$ line at 15:29:30 UT. The RB profile peaks at around $380~\rm{km~s^{-1}}$, which indicates a mass ejection moving towards Earth at the speed around $380~\rm{km~s^{-1}}$. Due to the blend at the red wing of Ne\,{\sc{vii}}\,$46.52~\rm{nm}$, we only conducted an RB asymmetry analysis for the O\,{\sc{iii}}\,$52.57~\rm{nm}$ and O\,{\sc{v}}\,$62.96~\rm{nm}$ lines. The velocities of the secondary component derived from the RB analysis are shown in Fig. \ref{fig:velos}(a).\\
\indent We then calculated the average LOS velocity of the ejecta by averaging all the velocities obtained from different lines using both methods at each time step during the period of 15:26:30--15:34:30 UT. The average LOS velocity {is} $423~\rm{km~s^{-1}}$ with a standard deviation of $40~\rm{km~s^{-1}}$ which we {take} as the uncertainty of the LOS velocity.\\

\subsection{{Full} velocity of the mass ejection}
\indent We have obtained the POS velocity $v_{\rm{POS}}$ from the viewpoint of \emph{STEREO-A} using time-distance diagrams in Sec. \ref{sec:velocity}. We have also estimated the LOS velocity $v_{\rm{LOS}}$ using \emph{SDO}/EVE observations around the Earth in Sec. \ref{sec:losvelo}. {The full velocity of the bulk motion of the ejecta} was then computed by combining these two velocities.\\
\indent {As the POS velocity is measured in the HPC system, we first converted it to the HEE system. We projected the POS velocity in the ecliptic plane (i.e. the X-Y plane of the HEE system) and perpendicular to the ecliptic plane (i.e. the Z-axis direction of the HEE system). {The deviation of \emph{STEREO-A} from the ecliptic plane, which is $0.075^\circ$ at $15:30~\rm{UT}$ on Oct. 28, 2021, can often be ignored as it is less than around $0.13^\circ$ throughout the STEREO era. Hence the Z-axis of the HPC system is in the ecliptic plane.} Because the Z-axis is perpendicular to the X-Y plane, the ecliptic plane containing the Z-axis is perpendicular to the X-Y plane of the HPC system. Therefore, the projections of the POS velocity $v_{\rm{POS}}$ in and perpendicular to the ecliptic plane are still in the X-Y plane of the HPC system. The angle between the $+$X-axis direction of the HPC system and the ecliptic plane (denoted as $\alpha$) and the angle between the POS velocity and $+$X-axis direction (i.e. $\theta$ in Eq. \eqref{theta}) can be directly added together to obtain the angle between the POS velocity and the ecliptic plane (denoted as $\theta_{\rm{POSP}}$). From the angle between the $+$Z-axis of the HEE system and $+$X-axis of the HPC system, $\alpha$ was derived to be around $1.6^\circ$, indicating that the $+$X-axis of the HPC system is north to the ecliptic plane. Therefore, $\theta_{\rm{POSP}}$ turned out to be around $-16.2^\circ$, which is negative meaning that the POS velocity is south to the ecliptic plane. The projected speeds of $v_{\rm{POS}}$ in and perpendicular to the ecliptic plane can be calculated by},
\begin{equation}
v_{\rm{POSE}}=v_{\rm{POS}}\cos{|\theta_{\rm{POSP}}|}
\end{equation}
\begin{equation}
v_{\rm{POSP}}=v_{\rm{POS}}\sin{|\theta_{\rm{POSP}}|}
\end{equation}
{where $v_{\rm{POSE}}$, around $564~\rm{km~s^{-1}}$, is in the ecliptic plane and $v_{\rm{POSP}}$, around $163~\rm{km~s^{-1}}$, is perpendicular to the ecliptic plane. It is also worth noting that the direction of $v_{\rm{POSE}}$ is perpendicular to the Sun-\emph{STEREO-A} line (i.e. the Z-axis of the HPC system).}\\
\indent {We combined $v_{\rm{LOS}}$, $v_{\rm{POSE}}$ and $v_{\rm{POSP}}$ to acquire the full velocity of the bulk motion of the ejecta. Firstly, we calculated the projected speed of full velocity of the ejecta in the ecliptic plane. A sketch for the geometry used in the calculation is displayed in Fig. \ref{fig:velos}(b). We denoted the projected speed in the ecliptic plane as $v_{\rm{E}}$ and the angle between the propagation direction of the ejecta and the Sun-Earth line in the ecliptic plane as $\theta_{\rm{E}}$. The angle between the Sun-\emph{STEREO-A} line and the Sun-Earth line measured in the ecliptic plane ($\theta_{\rm{STA}}$) is around $-37.5^\circ$. So we have,
\begin{equation}
\left\{
\begin{array}{ll}
v_{\rm{LOS}}=v_{\rm{E}}\cos\theta_{\rm{E}}\\ v_{\rm{POSE}}=v_{\rm{E}}\cos(90^\circ+\theta_{\rm{STA}}-\theta_{\rm{E}})
\end{array}
\right.
\label{ves}
\end{equation}
from which we derived,}
\begin{equation}
\begin{split}
&\frac{v_{\rm{LOS}}}{\cos\theta_{\rm{E}}}=\frac{v_{\rm{POSE}}}{\cos(90^\circ+\theta_{\rm{STA}}-\theta_{\rm{E}})}\\
\Longrightarrow &\theta_{\rm{E}}=\arctan \frac{v_{\rm{POSE}}/v_{\rm{LOS}}+\sin\theta_{\rm{STA}}}{\cos\theta_{\rm{STA}}}
\end{split}
\label{thetae}
\end{equation}
{The projected speed in the ecliptic plane ($v_{\rm{E}}$) is then obtained by Eq. \eqref{ves}. Secondly, we acquired the propagation direction and the full speed of the bulk motion of the ejecta by combing the velocities in and perpendicular to the ecliptic plane. The projected speed perpendicular to the ecliptic plane (denoted as $v_{\rm{P}}$) equals $v_{\rm{POSP}}$, thus the full velocity of the bulk motion of the ejecta can be calculated as,}
\begin{equation}
v=\sqrt{v_{\rm{E}}^2+v_{\rm{P}}^2},~v_{\rm{P}}=v_{\rm{POSP}}
\end{equation}
\begin{equation}
\theta_{\rm{P}}=\arctan{\frac{v_{\rm{POS}}\sin{\theta_{\rm{POSP}}}}{v_{\rm{E}}}}
\end{equation}
{where $v$ is the full speed of the ejecta and $\theta_{\rm{P}}$ is the angle between the true velocity vector and the ecliptic plane. {Negative $\theta_{\rm{P}}$ indicates that the full velocity is south to the ecliptic plane.} The direction of the bulk motion of the ejecta is described by $\theta_{\rm{E}}$ and $\theta_{\rm{P}}$. A 3D interactive version of the full velocity calculation is available with a snapshot shown in Fig. \ref{fig:velos}(c). All the velocity vectors are moved to the origin of the HEE system for plotting but it is worth noting that the ejecta originates from the AR which is located near the centeral meridian in the southern hemisphere of the Sun. The full velocity of the bulk motion of the ejecta was found to be around $596~\rm{km~s^{-1}}$ with a propagation direction of around $42.4^\circ$ west to the Sun-Earth line in the ecliptic plane and $16.0^\circ$ deviating to the south of the ecliptic plane.}\\
\indent {The above derivation of the full velocity becomes unusable when $|\theta_{\rm{STA}}|$ equals $90^\circ$. Under this condition, the POS velocity will lie in the X-Z plane of the HEE system and the solution of the full velocity is not unique. Also, the uncertainty in the full velocity estimation becomes fairly large when $|\theta_{\rm{STA}}|$ is close to $90^\circ$.}

\section{summary}\label{sec:conclusion}
\indent We have investigated signatures of a solar eruption on October 28, 2021 in the full-disk integrated spectra of He\,{\sc{i}}\,$58.43~\rm{nm}$, O\,{\sc{iii}}\,$52.57~\rm{nm}$, O\,{\sc{v}}\,$62.96~\rm{nm}$, O\,{\sc{vi}}\,$103.19~\rm{nm}$, Ne\,{\sc{vii}}\,$46.52~\rm{nm}$ and Ne\,{\sc{viii}}\,$77.03~\rm{nm}$ observed by \emph{SDO}/EVE. Intensities of all lines except O\,{\sc{iii}}\,$52.57~\rm{nm}$ experience hours-long dimmings after the flare, suggesting mass loss from the relatively lower solar atmosphere compared to normal CMEs. Obvious secondary components have been identified in the blue wings of O\,{\sc{iii}}\,$52.57~\rm{nm}$, O\,{\sc{v}}\,$62.96~\rm{nm}$, and Ne\,{\sc{vii}}\,$46.52~\rm{nm}$ during the impulsive phase and short after the flare peak, which is in accordance with the appearance of the mass ejection. A double Gaussian fitting and RB asymmetry analysis suggest that the mass ejection has a LOS velocity of around $423~\rm{km~s^{-1}}$ from the Earth viewpoint. With EUV imaging observations in the $304$~\AA~passband, we have also obtained the POS velocity of the ejecta from the \emph{STEREO-A} perspective. {Combing these two velocity components, the full velocity of the bulk motion of the ejecta has been found to be $\sim 596~\rm{km~s^{-1}}$ with an angle of $42.4^\circ$ west to the Sun-Earth line in the ecliptic plane and $16.0^\circ$ south to the ecliptic plane.}\\
\indent Our finding suggests that high-resolution spectroscopy at the EUV wavelengths has the potential to reveal the LOS velocities of CMEs. Combining the POS velocities inferred from EUV imaging observations, one could determine the true velocities of CMEs during their early propagation phases, which should significantly improve the accuracy of CME arrival time forecasting. The full-disk integrated spectra observed by EVE also provide a good opportunity to study the Sun as a remote star and improve our understanding of signals in the spectra from other stars. Our finding suggests that we may detect stellar CMEs through high-resolution spectroscopy, even though we cannot spatially resolve these features. This will be of great importance as stellar CMEs are the main drivers of exoplanetary space weather, which can significantly affect the habitability of exoplanets. Our future work will be focused on searching for the CME-related signals and estimating the physical parameters of CMEs through full-disk integrated spectra from both the Sun and other stars. {We may also combine CME models and theoretical spectra synthesis to examine the feasibility of detecting the blue-shifted secondary components in full-disk integrated spectra under different CME parameters and instrumental conditions.}\\

\section*{acknowledgments}
This work is supported by the National Key R\&D Program of China No. 2021YFA0718600, NSFC grants 11825301 \& 11790304. We would like to acknowledge the data use from \emph{GOES}, \emph{SDO}, and \emph{STEREO}/SECCHI. \emph{SDO} is a mission of NASA’s Living With a Star Program. Data from \emph{STEREO}/SECCHI are produced by the consortium of RAL (UK), NRL (USA), LMSAL (USA), GSFC (USA), MPS (Germany), CSL (Belgium), IOTA (France), and IAS (France). We appreciate the helpful discussion with P. C. Chamberlin, Li Feng and Beili Ying. We also thank the anonymous referee for his/her constructive comments and suggestions.

\bibliographystyle{aasjournal}
\bibliography{20211028}{}

\begin{thebibliography}{}
\expandafter\ifx\csname natexlab\endcsname\relax\def\natexlab#1{#1}\fi
\providecommand{\url}[1]{\href{#1}{#1}}
\providecommand{\dodoi}[1]{doi:~\href{http://doi.org/#1}{\nolinkurl{#1}}}
\providecommand{\doeprint}[1]{\href{http://ascl.net/#1}{\nolinkurl{http://ascl.net/#1}}}
\providecommand{\doarXiv}[1]{\href{https://arxiv.org/abs/#1}{\nolinkurl{https://arxiv.org/abs/#1}}}

\bibitem[{{Braga} {et~al.}(2017){Braga}, {Dal Lago}, {Echer}, {Stenborg}, \&
  {Rodrigues Souza de Mendon{\c{c}}a}}]{Braga2017}
{Braga}, C.~R., {Dal Lago}, A., {Echer}, E., {Stenborg}, G., \& {Rodrigues
  Souza de Mendon{\c{c}}a}, R. 2017, \apj, 842, 134,
  \dodoi{10.3847/1538-4357/aa755f}

\bibitem[{{Brown} {et~al.}(2016){Brown}, {Fletcher}, \& {Labrosse}}]{Brown2016}
{Brown}, S.~A., {Fletcher}, L., \& {Labrosse}, N. 2016, \aap, 596, A51,
  \dodoi{10.1051/0004-6361/201628390}

\bibitem[{{Brueckner} {et~al.}(1995){Brueckner}, {Howard}, {Koomen},
  {Korendyke}, {Michels}, {Moses}, {Socker}, {Dere}, {Lamy}, {Llebaria},
  {Bout}, {Schwenn}, {Simnett}, {Bedford}, \& {Eyles}}]{Brueckner1995}
{Brueckner}, G.~E., {Howard}, R.~A., {Koomen}, M.~J., {et~al.} 1995, \solphys,
  162, 357, \dodoi{10.1007/BF00733434}

\bibitem[{{Chamberlin}(2016)}]{Chamberlin2016}
{Chamberlin}, P.~C. 2016, \solphys, 291, 1665,
  \dodoi{10.1007/s11207-016-0931-0}

\bibitem[{{Cheng} {et~al.}(2019){Cheng}, {Wang}, {Liu}, {Zhou}, \&
  {Liu}}]{Cheng2019}
{Cheng}, Z., {Wang}, Y., {Liu}, R., {Zhou}, Z., \& {Liu}, K. 2019, \apj, 875,
  93, \dodoi{10.3847/1538-4357/ab0f2d}

\bibitem[{{De Pontieu} {et~al.}(2009){De Pontieu}, {McIntosh}, {Hansteen}, \&
  {Schrijver}}]{DePontieu2009}
{De Pontieu}, B., {McIntosh}, S.~W., {Hansteen}, V.~H., \& {Schrijver}, C.~J.
  2009, \apjl, 701, L1, \dodoi{10.1088/0004-637X/701/1/L1}

\bibitem[{{Del Zanna} {et~al.}(2021){Del Zanna}, {Dere}, {Young}, \&
  {Landi}}]{DelZanna2021}
{Del Zanna}, G., {Dere}, K.~P., {Young}, P.~R., \& {Landi}, E. 2021, \apj, 909,
  38, \dodoi{10.3847/1538-4357/abd8ce}

\bibitem[{{Dere} {et~al.}(1997){Dere}, {Landi}, {Mason}, {Monsignori Fossi}, \&
  {Young}}]{Dere1997}
{Dere}, K.~P., {Landi}, E., {Mason}, H.~E., {Monsignori Fossi}, B.~C., \&
  {Young}, P.~R. 1997, \aaps, 125, 149, \dodoi{10.1051/aas:1997368}

\bibitem[{{Echer} {et~al.}(2008){Echer}, {Gonzalez}, {Tsurutani}, \&
  {Gonzalez}}]{Echer2008}
{Echer}, E., {Gonzalez}, W.~D., {Tsurutani}, B.~T., \& {Gonzalez}, A.~L.~C.
  2008, Journal of Geophysical Research (Space Physics), 113, A05221,
  \dodoi{10.1029/2007JA012744}

\bibitem[{{Feng} {et~al.}(2013){Feng}, {Inhester}, \& {Mierla}}]{Feng2013}
{Feng}, L., {Inhester}, B., \& {Mierla}, M. 2013, \solphys, 282, 221,
  \dodoi{10.1007/s11207-012-0143-1}

\bibitem[{{Feng} {et~al.}(2012){Feng}, {Inhester}, {Wei}, {Gan}, {Zhang}, \&
  {Wang}}]{Feng2012}
{Feng}, L., {Inhester}, B., {Wei}, Y., {et~al.} 2012, \apj, 751, 18,
  \dodoi{10.1088/0004-637X/751/1/18}

\bibitem[{{Gonzalez} {et~al.}(1994){Gonzalez}, {Joselyn}, {Kamide}, {Kroehl},
  {Rostoker}, {Tsurutani}, \& {Vasyliunas}}]{Gonzalez1994}
{Gonzalez}, W.~D., {Joselyn}, J.~A., {Kamide}, Y., {et~al.} 1994, \jgr, 99,
  5771, \dodoi{10.1029/93JA02867}

\bibitem[{{Harra} {et~al.}(2016){Harra}, {Schrijver}, {Janvier}, {Toriumi},
  {Hudson}, {Matthews}, {Woods}, {Hara}, {Guedel}, {Kowalski}, {Osten},
  {Kusano}, \& {Lueftinger}}]{Harra2016}
{Harra}, L.~K., {Schrijver}, C.~J., {Janvier}, M., {et~al.} 2016, \solphys,
  291, 1761, \dodoi{10.1007/s11207-016-0923-0}

\bibitem[{Hou {et~al.}(2022)Hou, Tian, Wang, Zhang, Song, Zheng, Chen, Chen,
  Bai, Chen, He, Song, Zhang, Hu, Dun, Zong, Song, Xu, \& Tan}]{Hou2022}
Hou, Z., Tian, H., Wang, J.-S., {et~al.} 2022, The Astrophysical Journal, 928,
  98, \dodoi{10.3847/1538-4357/ac590d}

\bibitem[{{Howard} {et~al.}(2008){Howard}, {Moses}, {Vourlidas}, {Newmark},
  {Socker}, {Plunkett}, {Korendyke}, {Cook}, {Hurley}, {Davila}, {Thompson},
  {St Cyr}, {Mentzell}, {Mehalick}, {Lemen}, {Wuelser}, {Duncan}, {Tarbell},
  {Wolfson}, {Moore}, {Harrison}, {Waltham}, {Lang}, {Davis}, {Eyles},
  {Mapson-Menard}, {Simnett}, {Halain}, {Defise}, {Mazy}, {Rochus}, {Mercier},
  {Ravet}, {Delmotte}, {Auchere}, {Delaboudiniere}, {Bothmer}, {Deutsch},
  {Wang}, {Rich}, {Cooper}, {Stephens}, {Maahs}, {Baugh}, {McMullin}, \&
  {Carter}}]{Howard2008}
{Howard}, R.~A., {Moses}, J.~D., {Vourlidas}, A., {et~al.} 2008, \ssr, 136, 67,
  \dodoi{10.1007/s11214-008-9341-4}

\bibitem[{{Hudson} {et~al.}(2011){Hudson}, {Woods}, {Chamberlin}, {Fletcher},
  {Del Zanna}, {Didkovsky}, {Labrosse}, \& {Graham}}]{Hudson2011}
{Hudson}, H.~S., {Woods}, T.~N., {Chamberlin}, P.~C., {et~al.} 2011, \solphys,
  273, 69, \dodoi{10.1007/s11207-011-9862-y}

\bibitem[{{Inhester}(2006)}]{Inhester2006}
{Inhester}, B. 2006, arXiv e-prints, astro.
\newblock \doarXiv{astro-ph/0612649}

\bibitem[{{Kaiser} {et~al.}(2008){Kaiser}, {Kucera}, {Davila}, {St. Cyr},
  {Guhathakurta}, \& {Christian}}]{Kaiser2008}
{Kaiser}, M.~L., {Kucera}, T.~A., {Davila}, J.~M., {et~al.} 2008, \ssr, 136, 5,
  \dodoi{10.1007/s11214-007-9277-0}

\bibitem[{{Kim} {et~al.}(2008){Kim}, {Cho}, {Kim}, {Park}, {Moon}, {Yi}, {Lee},
  {Wang}, {Song}, \& {Dryer}}]{Kim2008}
{Kim}, R.~S., {Cho}, K.~S., {Kim}, K.~H., {et~al.} 2008, \apj, 677, 1378,
  \dodoi{10.1086/528928}

\bibitem[{{Kohl} {et~al.}(1995){Kohl}, {Esser}, {Gardner}, {Habbal},
  {Daigneau}, {Dennis}, {Nystrom}, {Panasyuk}, {Raymond}, {Smith}, {Strachan},
  {van Ballegooijen}, {Noci}, {Fineschi}, {Romoli}, {Ciaravella}, {Modigliani},
  {Huber}, {Antonucci}, {Benna}, {Giordano}, {Tondello}, {Nicolosi}, {Naletto},
  {Pernechele}, {Spadaro}, {Poletto}, {Livi}, {von der L{\"u}he}, {Geiss},
  {Timothy}, {Gloeckler}, {Allegra}, {Basile}, {Brusa}, {Wood}, {Siegmund},
  {Fowler}, {Fisher}, \& {Jhabvala}}]{Kohl1995}
{Kohl}, J.~L., {Esser}, R., {Gardner}, L.~D., {et~al.} 1995, \solphys, 162,
  313, \dodoi{10.1007/BF00733433}

\bibitem[{{Lemen} {et~al.}(2012){Lemen}, {Title}, {Akin}, {Boerner}, {Chou},
  {Drake}, {Duncan}, {Edwards}, {Friedlaender}, {Heyman}, {Hurlburt}, {Katz},
  {Kushner}, {Levay}, {Lindgren}, {Mathur}, {McFeaters}, {Mitchell}, {Rehse},
  {Schrijver}, {Springer}, {Stern}, {Tarbell}, {Wuelser}, {Wolfson}, {Yanari},
  {Bookbinder}, {Cheimets}, {Caldwell}, {Deluca}, {Gates}, {Golub}, {Park},
  {Podgorski}, {Bush}, {Scherrer}, {Gummin}, {Smith}, {Auker}, {Jerram},
  {Pool}, {Soufli}, {Windt}, {Beardsley}, {Clapp}, {Lang}, \&
  {Waltham}}]{Lemen2012}
{Lemen}, J.~R., {Title}, A.~M., {Akin}, D.~J., {et~al.} 2012, \solphys, 275,
  17, \dodoi{10.1007/s11207-011-9776-8}

\bibitem[{{Liewer} {et~al.}(2011){Liewer}, {Hall}, {Howard}, {De Jong},
  {Thompson}, \& {Thernisien}}]{Liewer2011}
{Liewer}, P.~C., {Hall}, J.~R., {Howard}, R.~A., {et~al.} 2011, Journal of
  Atmospheric and Solar-Terrestrial Physics, 73, 1173,
  \dodoi{10.1016/j.jastp.2010.09.004}

\bibitem[{{Mason} {et~al.}(2014){Mason}, {Woods}, {Caspi}, {Thompson}, \&
  {Hock}}]{Mason2014}
{Mason}, J.~P., {Woods}, T.~N., {Caspi}, A., {Thompson}, B.~J., \& {Hock},
  R.~A. 2014, \apj, 789, 61, \dodoi{10.1088/0004-637X/789/1/61}

\bibitem[{{Mason} {et~al.}(2016){Mason}, {Woods}, {Webb}, {Thompson},
  {Colaninno}, \& {Vourlidas}}]{Mason2016}
{Mason}, J.~P., {Woods}, T.~N., {Webb}, D.~F., {et~al.} 2016, \apj, 830, 20,
  \dodoi{10.3847/0004-637X/830/1/20}

\bibitem[{{Mishra} \& {Srivastava}(2013)}]{Mishra2013}
{Mishra}, W., \& {Srivastava}, N. 2013, \apj, 772, 70,
  \dodoi{10.1088/0004-637X/772/1/70}

\bibitem[{{Moon} {et~al.}(2005){Moon}, {Cho}, {Dryer}, {Kim}, {Bong}, {Chae},
  \& {Park}}]{Moon2005}
{Moon}, Y.~J., {Cho}, K.~S., {Dryer}, M., {et~al.} 2005, \apj, 624, 414,
  \dodoi{10.1086/428880}

\bibitem[{{Moran} \& {Davila}(2004)}]{Moran2004}
{Moran}, T.~G., \& {Davila}, J.~M. 2004, Science, 305, 66,
  \dodoi{10.1126/science.1098937}

\bibitem[{{Pesnell} {et~al.}(2012){Pesnell}, {Thompson}, \&
  {Chamberlin}}]{Pesnell2012}
{Pesnell}, W.~D., {Thompson}, B.~J., \& {Chamberlin}, P.~C. 2012, \solphys,
  275, 3, \dodoi{10.1007/s11207-011-9841-3}

\bibitem[{{Susino} {et~al.}(2014){Susino}, {Bemporad}, \& {Dolei}}]{Susino2014}
{Susino}, R., {Bemporad}, A., \& {Dolei}, S. 2014, \apj, 790, 25,
  \dodoi{10.1088/0004-637X/790/1/25}

\bibitem[{{Susino} {et~al.}(2013){Susino}, {Bemporad}, {Dolei}, \&
  {Vourlidas}}]{Susino2013}
{Susino}, R., {Bemporad}, A., {Dolei}, S., \& {Vourlidas}, A. 2013, Advances in
  Space Research, 52, 957, \dodoi{10.1016/j.asr.2013.05.017}

\bibitem[{{Thompson}(2006)}]{Thompson2006}
{Thompson}, W.~T. 2006, \aap, 449, 791, \dodoi{10.1051/0004-6361:20054262}

\bibitem[{{Thompson}(2009)}]{Thompson2009}
---. 2009, \icarus, 200, 351, \dodoi{10.1016/j.icarus.2008.12.011}

\bibitem[{{Tian} {et~al.}(2011){Tian}, {McIntosh}, {De Pontieu},
  {Mart{\'\i}nez-Sykora}, {Sechler}, \& {Wang}}]{Tian2011}
{Tian}, H., {McIntosh}, S.~W., {De Pontieu}, B., {et~al.} 2011, \apj, 738, 18,
  \dodoi{10.1088/0004-637X/738/1/18}

\bibitem[{{Tian} {et~al.}(2012){Tian}, {McIntosh}, {Xia}, {He}, \&
  {Wang}}]{Tian2012}
{Tian}, H., {McIntosh}, S.~W., {Xia}, L., {He}, J., \& {Wang}, X. 2012, \apj,
  748, 106, \dodoi{10.1088/0004-637X/748/2/106}

\bibitem[{{Tian} {et~al.}(2013){Tian}, {Tomczyk}, {McIntosh}, {Bethge}, {de
  Toma}, \& {Gibson}}]{Tian2013}
{Tian}, H., {Tomczyk}, S., {McIntosh}, S.~W., {et~al.} 2013, \solphys, 288,
  637, \dodoi{10.1007/s11207-013-0317-5}

\bibitem[{{Woods} {et~al.}(2012){Woods}, {Eparvier}, {Hock}, {Jones},
  {Woodraska}, {Judge}, {Didkovsky}, {Lean}, {Mariska}, {Warren}, {McMullin},
  {Chamberlin}, {Berthiaume}, {Bailey}, {Fuller-Rowell}, {Sojka}, {Tobiska}, \&
  {Viereck}}]{Woods2012}
{Woods}, T.~N., {Eparvier}, F.~G., {Hock}, R., {et~al.} 2012, \solphys, 275,
  115, \dodoi{10.1007/s11207-009-9487-6}

\bibitem[{{Wuelser} {et~al.}(2004){Wuelser}, {Lemen}, {Tarbell}, {Wolfson},
  {Cannon}, {Carpenter}, {Duncan}, {Gradwohl}, {Meyer}, {Moore}, {Navarro},
  {Pearson}, {Rossi}, {Springer}, {Howard}, {Moses}, {Newmark},
  {Delaboudiniere}, {Artzner}, {Auchere}, {Bougnet}, {Bouyries}, {Bridou},
  {Clotaire}, {Colas}, {Delmotte}, {Jerome}, {Lamare}, {Mercier}, {Mullot},
  {Ravet}, {Song}, {Bothmer}, \& {Deutsch}}]{Wulser2004}
{Wuelser}, J.-P., {Lemen}, J.~R., {Tarbell}, T.~D., {et~al.} 2004, in Society
  of Photo-Optical Instrumentation Engineers (SPIE) Conference Series, Vol.
  5171, Telescopes and Instrumentation for Solar Astrophysics, ed.
  S.~{Fineschi} \& M.~A. {Gummin}, 111--122, \dodoi{10.1117/12.506877}

\bibitem[{{Yang} {et~al.}(2022){Yang}, {Tian}, {Bai}, {Chen}, {Guo}, {Zhu},
  {Cheng}, {Gao}, {Xu}, {Chen}, \& {Zhang}}]{Yang2022}
{Yang}, Z., {Tian}, H., {Bai}, X.-Y., {et~al.} 2022, ApJS, in press

\bibitem[{{Ying} {et~al.}(2022){Ying}, {Feng}, {Inhester}, {Mierla}, {Gan},
  {Lu}, \& {Li}}]{Ying2022}
{Ying}, B., {Feng}, L., {Inhester}, B., {et~al.} 2022, \aap, 660, A23,
  \dodoi{10.1051/0004-6361/202142797}

\end{thebibliography}

\end{CJK*}
\end{document}